\documentclass[twocolumn,preprintnumbers,amsmath,amssymb]{revtex4}

\usepackage{graphicx}
\usepackage{dcolumn}
\usepackage{bm}

\begin{document}

\title{Exploiting Locally Imposed Anisotropies in (Ga,Mn)As: a Non-volatile Memory Device}

\author{K. Pappert}%
\author{S. H\"{u}mpfner}
\author{C. Gould}%
\author{J. Wenisch}%
\author{K. Brunner}%
\author{G. Schmidt}%
\author{L.W. Molenkamp}%

\affiliation{%
Physikalisches Institut (EP3), Universit\"{a}t W\"{u}rzburg, Am
Hubland, D-97074 W\"{u}rzburg, Germany }%

\date{\today}
\begin{abstract}
Progress in (Ga,Mn)As lithography has recently allowed us to
realize structures where unique magnetic anisotropy properties can
be imposed locally in various regions of a given device. We make
use of this technology to fabricate a device in which we study
transport through a constriction separating two regions whose
magnetization direction differs by $90^\circ$. We find that the
resistance of the constriction depends on the flow of the magnetic
field lines in the constriction region and demonstrate that such a
structure constitutes a non-volatile memory device.
\end{abstract}

\maketitle

(Ga,Mn)As can be regarded as a prototypical material for
investigating potential device applications of ferromagnetic
semiconductors. The spin-orbit mediated coupling of magnetic and
semiconductor properties in this material gives rise to many novel
transport-related phenomena which can be harnessed for device
applications. Previously reported device concepts include strong
anisotropic magnetoresistance (AMR), in-plane Hall effect
\cite{Roukes}, tunneling anisotropic magnetoresistance (TAMR)
\cite{GouldTAMR,ChrisTAMR,KatrinTAMR} and Coulomb blockade
AMR\cite{CBAMR}. These previous demonstrations have been based on
structures which have the same magnetic properties, inherited from
the unstructured (Ga,Mn)As layer, throughout the device.
Improvement in lithographic capabilities \cite{nanobars} has
recently allowed for the first time the production of structures
where distinct anisotropies are imposed locally to various
functional elements of the same device by overwriting the parent
layer anisotropy. This greatly enhances the scope of possible
device paradigms open to investigation as it allows for devices
where the functional element involves transport between regions
with different magnetic anisotropy properties.

In this letter we present the first such device. It is comprised
of two (Ga,Mn)As nanobars, oriented perpendicular to each other,
and with each nanobar exhibiting strong uniaxial magnetic
anisotropy. These two nanobars are electrically connected through
a constriction whose resistance is determined by the relative
magnetization states of the nanobars. We show that the anisotropic
magnetoresistance effect yields different constriction resistances
depending on the relative orientation of the two
nanobar-magnetization vectors. The structure can thus be viewed as
the basis of a ferromagnetic semiconductor memory device that
operates in the non-volatile regime.

\begin{figure}[b] \includegraphics{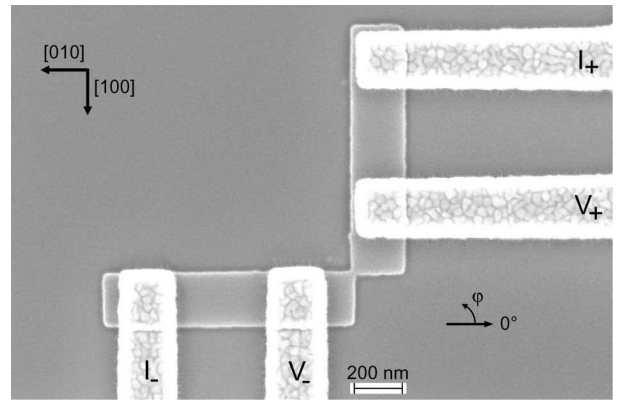}
\caption{\label{f1} SEM photograph of the device identifying the
orientation of the nanobars with respect to the crystal
directions, and the definition of the current (I+,I-) and voltage
(V+,V-) leads and the writing angle $\varphi$}
\end{figure}

For the device, we use a 20 nm thick (Ga,Mn)As layer grown on a
GaAs substrate \cite{growth} by low-temperature molecular beam
epitaxy. Using an electron-beam lithography(EBL) defined Ti-mask
and chemically assisted ion beam etching (CAIBE) this layer is
patterned into several pairs of coupled nanobars\cite{nanobars} as
shown in the SEM micrograph in Fig.~\ref{f1}. Ti/Au contacts are
defined in another EBL-step through metal evaporation and
lift-off, yielding resistance area products of $\sim1\mu\Omega
$cm$^2$. The bars are circa 200 nm wide and 1$\mu$m long and
oriented along the [100] and [010] crystal direction,
respectively. They form a $90^\circ$-angle and touch each other in
one corner, where a constriction with a width of some tens of nm
is formed.

\begin{figure}[t]
    \centering
        \includegraphics[width=8.5cm]{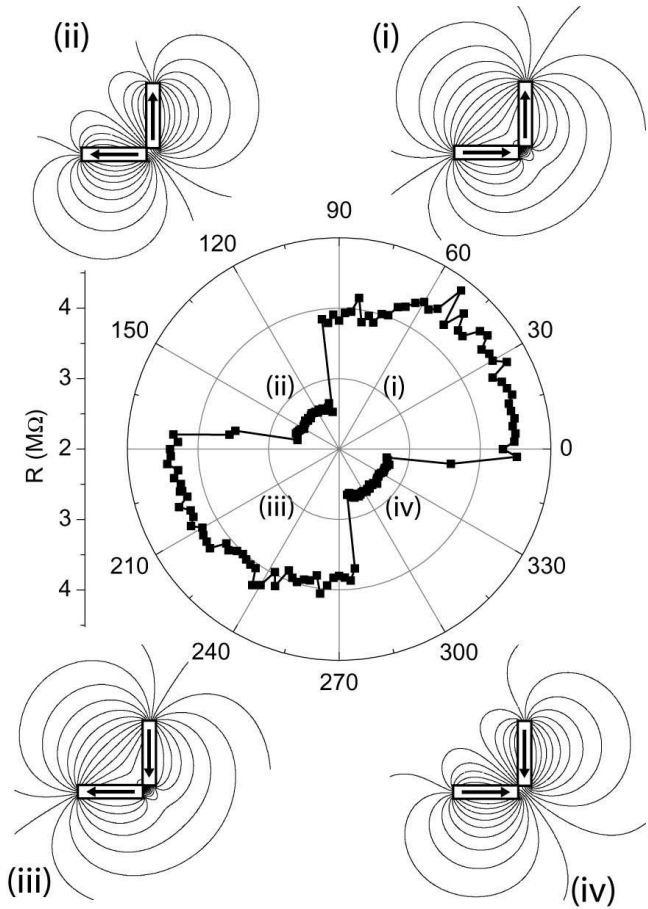}
    \caption{Polar plot showing the results of a "write-read" experiment.
     The state of the device is written by applying a magnetic field of 300 mT in the $\varphi$ direction.
     This field is then swept back to zero, and the resistance of the device is measured. The insets sketch the magnetic configuration of the device in each quadrant and the
     corresponding field line patterns\cite{Vizimag}.}
    \label{F2}
\end{figure}

Transport measurements are carried out at 4 K in a magnetocryostat
fitted with a vector field magnet that allows the application of a
magnetic field of up to 300 mT in any direction. The sample state is
first "written" by an in-plane magnetic field of 300 mT along a
writing angle $\varphi$ (as defined in Fig.~\ref{f1}). The field is
then slowly swept back to zero while ensuring that the magnetic
field vector never deviates from the $\varphi$-direction. We measure
the four-terminal resistance of the constriction in the resulting
remanent state by applying a voltage $V_{b}$ to the current leads
($I_{+}$ and $I_{-}$), and recording both the voltage drop between
contacts $V_{+}$ and $V_{-}$ and the current that is flowing from
$I_{+}$ to $I_{-}$ (Fig.~\ref{f1}). The polar plot of Fig.~\ref{F2}
shows the constriction resistance of the remanent magnetization
state as a function of the writing angle $\varphi$. The resistance,
which is dominated by the constriction, has a higher value upon
writing the sample in the (extended) first quadrant
($-3^\circ\leq\varphi<98^\circ$) and a lower value upon writing in
the (shrunken) second quadrant ($98^\circ<\varphi<167^\circ$). As a
whole the plot is point-symmetric with respect to the origin.

\begin{figure}[t]
    \centering
        \includegraphics[width=8.5cm]{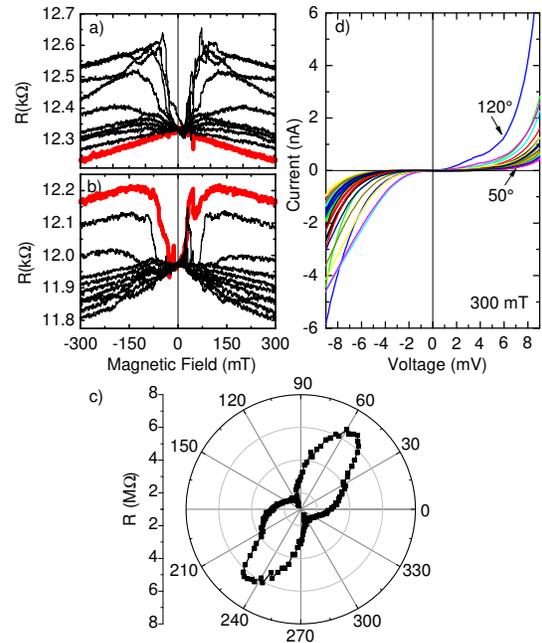}
    \caption{Magnetoresistance measurements on the $0^\circ$(a) and $90^\circ-$nanobar(b) confirming that each exhibits a
    strong uniaxial magnetic anisotropy along its long axis. Field sweeps from -300 to 300 mT along
    $0^\circ$(thick line)$\ldots 90^\circ$ in $10^\circ$ steps. c) Polar plot showing the resistance of the
    constriction in a field of 300 mT at various angles.}
    \label{F3}
\end{figure}

To explain these results, we first examine the behavior of the
individual nanobars. They are patterned on the sub-micron scale to
make use of anisotropic strain relaxation, which in turn causes a
uniaxial magnetic anisotropy that is strong enough to overwrite
the intrinsic anisotropy of the (Ga,Mn)As layer\cite{nanobars}. We
therefore expect each nanobar to show a uniaxial magnetic
anisotropy with a magnetic easy axis along the respective long
axis of each of the nanobars. That this is true also for coupled
nanobars is confirmed in Fig.~\ref{F3}, which shows two terminal
magnetoresistance scans, performed separately on the
$0^\circ$-nanobar (Fig.~\ref{F3}a) and the $90^\circ$-nanobar
(Fig.~\ref{F3}b) pictured in Fig.~\ref{f1}. The plots show field
sweeps from -300 to +300 mT for various in-plane field directions
$\varphi$ between $0^\circ$ and $90^\circ$. Metallic (Ga,Mn)As
exhibits a higher resistance value when the magnetization
\textbf{M} is perpendicular to the current \textbf{J}, than when
\textbf{M} is parallel to \textbf{J} (this is the AMR effect
\cite{Jan},\cite{AMRGaMnAs}). When the field \textbf{H} is swept
along $0^\circ$ (thick line in Fig.~\ref{F3}a), the resistance of
the $0^\circ$-nanobar remains in the low state \cite{NOTE},
indicating that \textbf{M} is parallel to \textbf{J} throughout
the entire magnetic field range. All the other MR-scans start at a
higher resistance value and merge into the low resistance curve at
zero field, indicating that \textbf{M}, which is almost parallel
to \textbf{H} at high fields, relaxes towards the $0^\circ$
uniaxial easy axis as the field is decreased. Analogously, the
uniaxial easy axis of the $90^\circ$-nanobar is along $90^\circ$
(Fig.~\ref{F3}b). Consequently, the $90^\circ$-MR-scan is a flat
low resistance curve. During the $0^\circ$-scan (thick line) the
magnetization relaxes from parallel to the field (high resistance)
towards the easy axis along the bar (low resistance) at zero
field.

Given that both bars show a uniaxial magnetic easy axis along
their respective long axis, the structure has four possible
magnetic states at zero magnetic field as sketched in
Fig.~\ref{F2}. In sectors (i) and (iii) the nanobars are
magnetized "in series",i.e. the magnetization vectors meet in a
configuration which we will call \emph{head-to-tail}. In (ii) and
(iv) on the other hand, both magnetization vectors point away from
(\emph{tail-to-tail}) or towards (\emph{head-to-head}) the
constriction. When the sample is magnetized along a given
direction at 300 mT, the magnetization of both bars is almost
parallel to the magnetic field. As the field is then lowered to
zero, the magnetization of each nanobar relaxes to the respective
nanobar easy axis, selecting the direction which is closest to the
writing angle $\varphi$. For a nanobar along $0^\circ$ this means,
assuming no interaction between the bars, that \textbf{M} relaxes
to $0^\circ$ upon writing the bar along any angle between
$+90^\circ$ and $-90^\circ$; otherwise \textbf{M} relaxes to
$180^\circ$. If the bars in our device were non-interacting, one
would thus expect the magnetization configuration in each quadrant
to be as depicted in the sketches of Fig.~\ref{F2}, with each
quadrant accounting for exactly one fourth of the total plot.

The deviation from this behavior in the actual device is due to
magnetostatic interactions between the two bars, which cause a
preference for head-to-tail configurations. A simple magnetostatic
calculation shows that the repulsive field felt by the tip of one
bar due to being near the wrong pole of the other bar is of the
order of 2 mT, which is $\sim5\%$ of the uniaxial anisotropy
field. The energy density of this field is thus strong enough to
overcome a small part of the energy barrier against rotation
towards the opposite magnetization direction, which corresponds to
an angle of $\sim 3^\circ$. The head-to-tail quadrants thus
increase commensurably.

Magnetic field line patterns for the four magnetization
configurations were calculated (sketches in Fig.~\ref{F2}i-iv)
using a simple bar magnet model. The field lines are close to
parallel to the current in the head-to-tail configuration
(Fig.~\ref{F2}i and iii).
In the tail-to-tail and the head-to-head
configuration (Fig.~\ref{F2}ii and iv) the field lines are
approximately perpendicular to the current.

\begin{figure}[t]
    \centering
        \includegraphics[width=9.0cm]{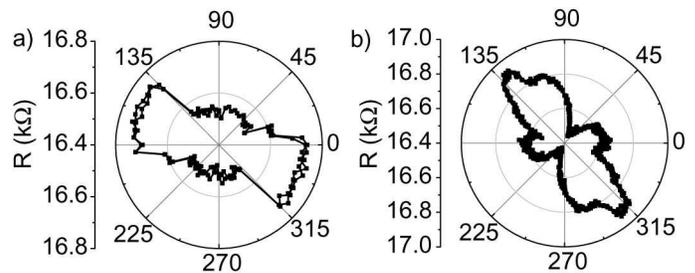}
    \caption{
    (a)Results of a write-read experiment as in Fig.~\ref{F2}, for
    a device with a wider constriction, which exhibits metallic transport behavior
and (b)constriction resistance in a rotating 300 mT external
magnetic field. }
    \label{F4}
\end{figure}

Having understood the magnetic configuration of the device in the
write-read experiment of Fig.~\ref{F2}, we now turn to an
explanation of why these should lead to two very distinct
resistance states. The above magnetostatic arguments and internal
fields, in connection with the AMR coefficient for metallic
(Ga,Mn)As can explain a few percent resistance difference
\cite{AMRGaMnAs,Roukes} between the head-to-tail and the
head-to-head configuration, much smaller and of a different sign
than the effect in Fig.~\ref{F2}. We have actually observed such a
small AMR related effect in a similar structure, which has a wider
constriction (100 times lower constriction resistance).
Fig.~\ref{F4}a shows similar data on this low resistance sample as
Fig.~\ref{F2} for the high resistance sample. It is immediately
obvious from Fig.~\ref{F4}a that this sample shows the same
remanent magnetization configurations as the device in
Fig.~\ref{F2}. However, the effect is much smaller and of the
opposite sign: where Fig.~\ref{F2} exhibits a high resistance
state, Fig.~\ref{F4}a shows a low state, and {\em vice versa}.

We ascribe the difference in behavior between Figs.~\ref{F2}a and
~\ref{F4}a to the occurrence of depletion in the constriction in
the sample of Fig.~\ref{F2}a, which drives the transport (in the
critical constriction region) into the hopping
regime\cite{ChrisNanoconstr}. At the same time, we suggest that in
the hopping regime the AMR coefficient {\em changes sign}, leading
to the observed changes in magnetoresistance. Important evidence
for this claim comes from the angle-dependent magnetoresistance
behavior of the samples at a field of 300 mT, strong enough to
force the magnetization close to parallel to the external field.
This data is given in Fig.~\ref{F3}c for the high-resistance, and
in Fig.~\ref{F4}b for the low-resistance sample.

The low-resistance device exhibits typical AMR behavior as
expected for metallic (Ga,Mn)As: Fig.~\ref{F4}b shows that the
resistance is lowest when \textbf{M} is forced parallel to the
current through the constriction ($\varphi \sim ~45^\circ$) and
ca. 3 \% higher for \textbf{M}$\bot$\textbf{J}. In contrast, the
high-resistance constriction of the device in Fig.~\ref{F2} shows
a huge and inverted AMR signal, as can be seen in Fig.~\ref{F3}c.
The resistance at $\varphi \sim ~45^\circ$, where
\textbf{M}$\|$\textbf{J}, is more than 5 times larger than for
\textbf{M}$\bot$\textbf{J}.

This is actually not the first observation of an inverted AMR
signal; the same effect has recently been reported in thin
(Ga,Mn)As devices \cite{CBAMR,Andrew} in which the transport is in
the hopping regime. This situation is similar to our
high-resistance device, where from the resistance one already can
infer that the constriction acts as a tunnel barrier. Actual
evidence for tunneling transport comes from the current-voltage
characteristics of the high-resistance constriction, shown in
Fig.~\ref{F3}d, which were taken at 300 mT at different field
directions $\varphi$. The I-V's are clearly non-linear, with the
nonlinearity depending on the magnetization direction. Fields
aligning \textbf{M} along $\sim120^\circ$ cause the strongest and
along $50^\circ$ the smallest non-linearity of the IV-curve.

The strong dependence of the IV-characteristic and the resistance
on the magnetization direction are characteristic of transport
going through a metal-insulator transition (MIT) from the
diffusive into the hopping regime depending on the angle of the
magnetization, similar to what we have previously observed in a
TAMR device \cite{KatrinTAMR}. Such a MIT occurs in partly
depleted samples due to the wave-function geometry change
depending on the magnetization direction. The localized hole
wave-function has an oblate shape with the smaller axis pointing
in the magnetization direction (\cite{wavefunction}). Consider the
overlap of such oblate shapes statistically distributed with
respect to the direction of the current in connection with the
Thouless localization criterion. The wavefunction overlap is much
smaller when the sample is magnetized parallel to the current,
than for $\textbf{M}\bot\textbf{J}$, suppressing hopping transport
through the depleted constriction region. This implies a
magnetoresistance behavior that is exactly the inverse of that
expected for the metallic regime and explains the increased
resistance value in both the high field measurements
(Fig.~\ref{F3}c along $\sim 45^\circ$) and the write-read
experiment (Fig.~\ref{F2} 1st quadrant).

We thus believe that our observations can be fully explained by
the internal magnetic fields and the AMR coefficient as applicable
to the transport regime in the constriction. A further candidate
to explain our observations could be the presence of a domain wall
(DW) between differently magnetized regions of the device in the
head-to-head and tail-to-tail configuration, which would be absent
in the head-to-tail configurations. However, since the
constriction is long and the DW would not be strongly
geometrically confined, one anticipates only a very low DW
resistance in these samples \cite{ChrisNanoconstr,Bruno}. This is
confirmed by a comparison of Fig.~\ref{F2} with Fig.~\ref{F3}c:
The resistance values of both remanent states in Fig.~\ref{F2} are
in between the extreme resistance values of the homogeneously
magnetized sample. The DW contribution \cite{ChrisNanoconstr} to
the constriction resistance can in the present sample thus only be
a minor effect on the resistance of the remanent state and does
not explain the different resistance levels in Fig.~\ref{F2}.
\cite{notegeorg}. In the remanent state the resistance of the
head-to-head configuration, including a possible DW contribution,
is lower than the resistance of the head-to-tail configuration. We
can thus exclude the DW as the origin of the two resistance states
observed in Fig.~\ref{F2}.

In conclusion we have shown that locally imposed magnetic
anisotropies in different regions of one ferromagnetic semiconductor
device allow for novel device designs. We consider the
perpendicularly magnetized nanobars discussed in this paper as a
first demonstration of the type of devices that can be fabricated
using this approach, it is certainly not difficult to conceive of
further concepts in this direction. In addition, the work presented
here has highlighted the difference in AMR behavior between metallic
and hopping transport in (Ga,Mn)As, which again should prove useful
in device design.

\begin{acknowledgments}
The authors thank M. Sawicki and M.J. Schmidt for useful discussions
and V. Hock and T. Borzenko for help in sample fabrication. We
acknowledge financial support from the EU (NANOSPIN FP6-IST-015728)
and the German DFG (BR1960/2-2).

\end{acknowledgments}


\begin{thebibliography}{99}
\bibitem{notegeorg} In a separate experiment we have prepared a DW pinned at the
smallest part of our constriction. This was accomplished using
well defined field sequences and will be published elsewhere. In
this case we observe a large positive DW contribution to the
device resistance (as reported in \cite{ChrisNanoconstr}).
\end{thebibliography}
\end{document}